\documentclass[12pt]{iopart}
\usepackage{iopams,graphicx,url}
\usepackage[hidelinks]{hyperref}

\newcommand\meshcells{700}
\newcommand\simtime{36}

\newcommand{\electron}{\ensuremath{\textrm{e}}}
\newcommand{\cfl}{\ensuremath{\textrm{cfl}}}
\newcommand{\diffusion}{\ensuremath{D}}
\newcommand{\bme}{\ensuremath{{\mathbf{E}}}}
\newcommand{\bmv}{\ensuremath{{\mathbf{v}}}}
\newcommand{\bmi}{\ensuremath{{\mathbf{i}}}}
\newcommand{\cm}{\ensuremath{{\textrm{cm}}}}
\newcommand{\um}{\ensuremath{{\mu\textrm{m}}}}
\newcommand{\m}{\ensuremath{{\textrm{m}}}}
\newcommand{\kV}{\ensuremath{{\textrm{kV}}}}
\def\ns{\ensuremath{{\textrm{ns}}}}
\newcommand{\fs}{\ensuremath{{\textrm{fs}}}}

\begin{document}
\title{Adaptive multiscale methods for 3D streamer discharges in air}
\author{Robert Marskar}
\address{SINTEF Energy Research, Sem S\ae lands vei 11, 7034 Trondheim, Norway.}
\ead{robert.marskar@sintef.no}
\date{\today}
\begin{abstract}
  We discuss spatially and temporally adaptive implicit-explicit (IMEX) methods for parallel simulations of three-dimensional fluid streamer discharges in atmospheric air. We examine strategies for advancing the fluid equations and elliptic transport equations (e.g. Poisson) with different time steps, synchronizing them on a global physical time scale which is taken to be proportional to the dielectric relaxation time. The use of a longer time step for the electric field leads to numerical errors that can be diagnosed, and we quantify the conditions where this simplification is valid. Likewise, using a three-term Helmholtz model for radiative transport, the same error diagnostics show that the radiative transport equations do not need to be resolved on time scales finer than the dielectric relaxation time. Elliptic equations are bottlenecks for most streamer simulation codes, and the results presented here potentially provide computational savings. Finally, a computational example of 3D branching streamers in a needle-plane geometry that uses up to \meshcells~million grid cells is presented.
\end{abstract}
\maketitle

\section{Introduction}
Streamer discharges are fast filamentary transients that evolve due to self-enhanced electric fields at their tips. Streamers are the natural precursors of sparks, lightning, and sprites, and have found use in sterilization of polluted gases and breakup of molecules, plasma assisted combustion, and control of airflow in the boundary layer of airplane wings \cite{Boeuf2005, Moreau2007, Soloviev2009, 1167639, 18870, 4504897, 55956, Nair2004, Grymonpre2001, Ebert2006,Bogaerts2002}. The spatial scales for streamer discharges span several orders of magnitude. Firstly, a streamer is essentially a non-thermal plasma filament surrounded by a space charge layer with a thickness one to two orders of magnitude thinner than the filament diameter. Secondly, the length of the filament can be much longer than its thickness, leading to a numerical problem with widely different spatial scales. On the one hand, a fine numerical resolution is required for resolving the space charge layer. On the other, a large computational domain is required in order to facilitate the propagation of the streamer. Due to the nonlinearity of streamers, numerical solutions can usually not be obtained on coarse grids.

Streamers can be described by using either fluid or kinetic approaches, or a combination of them \cite{Li2010, Li2012}. In a fluid approach, electrons and ions are evolved according to their fluid moments (usually truncated to first order) by using tabulated or analytic transport data. Kinetic approaches approximate the phase space distribution function by evolving computational particles, using cross-sectional collision data as input parameters. The kinetic approach is computationally far more exhaustive, and includes far more physics. Fluid approximations are more common on larger scales, and this is the description that we use in this paper. 

It is generally believed that spatially adaptive methods are ideally suited - and maybe even necessary - for large scale simulations of 3D streamers. The reason for this is that a streamer represents a dynamically evolving structure with possible stochastic behavior, restricting the use static grids to cases where the streamer path is a priori known, or small-scale cases where the entire domain can be resolved at the finest spatial resolution. Because of these difficulties, 3D simulations of streamers are rare \cite{Hallac2003, Pancheshnyi2008, Luque2008, Papageorgiou2011, Nijdam2016a, Teunissen2017, Teunissen2017, Plewa2018, Marskar2018}. In addition to multiple spatial scales, various characteristic time scales also exist for streamer discharges. For example, the avalanche-to-streamer time describes the time it takes for seed electrons to reach a critical size through impact ionization, whereas the dielectric relaxation time describes the local rate-of-change of the electric field due to motion of charge carriers. The relaxation time can be expressed in terms of the electric field $\mathbf{E}$ and the electric current density $\mathbf{J}$ as
\begin{equation}
  \Delta t_{\mathbf{E}} = \frac{\epsilon_0\left|\mathbf{E}\right|}{\left|\mathbf{J}\right|},
\end{equation}
which follows from a first-order trunction of Ampere's law. This time scale is an important one for streamer simulations. Note that $\Delta t_\bme$ is a \emph{physical} time scale that does not reference the grid resolution. 

Computer codes have restrictions on which time step sizes can be used to evolve the fluid equations. For example, explicit methods for advection obey a Courant-Friedrichs-Lewy (CFL) time step restriction $\Delta t_{\cfl} = \Delta x/|\mathbf{v}|$ where $\Delta x$ is the spatial resolution and $\mathbf{v}$ is the advective velocity.  For parabolic equations, such as diffusion equations, the time step restriction on explicit methods is $\Delta t_{\diffusion} \sim \Delta x^2$, which quickly becomes even stricter than $\Delta t_\cfl$ when diffusion is relevant. Furthermore, although not required for stability, one may derived a fourth time step constraint $\Delta t_{S} = \frac{S}{n}$, where $S$ is a source term and $n$ is a species density, which describes a characteristic time scale for growth due to ionization. For multiple components, the above time step sizes also minimized over species according to the worst offender, which is usually the electrons. On the other hand, kinetic models like Particle-In-Cell (PIC) choose their time steps differently, being restricted by the inverse plasma frequency and the inverse collision frequencies. 

For streamer simulations, the above mentioned time scales are widely varying. Fully implicit methods are likely to be the most robust, but it is not yet clear how these should be implemented in the multiphysics environment of streamers, particularly if high-order shock-capturing methods on adaptive grids are involved for the convection part. For that reason, implicit-explicit (IMEX) methods are more attractive, allowing one to treat restrictive time scales with implicit methods and use explicit methods on others. Previous experience with large scale 3D simulations \cite{Marskar2018} show that such methods are often CFL bound on the time step, leading us to speculate that run-time performance can be improved through a better segregation between physical and numerical time scales. The rationale behind this idea is that it should be unnecessary to perform electric field updates when $\Delta t_{\cfl} \ll \Delta t_{\bme}$, since the correction to the $\bme$-field would then be miniscule. Similar ideas have been adopted in particle codes \cite{Fierro2014, Teunissen2016}. Likewise, diffusive models approximations of the radiative transport equation reduce to Helmholtz equations, and it is neither not clear how often these need to be updated. 

This paper quantifies the prospects of advancing fluid and elliptic equations using different time steps. Potentially, such techniques may lead to speedup (and flexibility) for fluid streamer simulation by elimination of extranous elliptic solves at very fine time scales, which are present for consistent schemes. The outline of this paper is as follows: In Sec.~\ref{sec:theory} we present a time stepping scheme that segregates the numerical and physical time scales. The numerical error of this scheme is then quantified in Sec.~\ref{sec:2D} by means of two-dimensional simulation experiments. Thus, we partially answer the question ''How often do we need to update the electric field?'', which is of interest to both the kinetic and fluid modeling parts of the plasma community. Finally, we present a high-performance computing example in Sec.~\ref{sec:hpc_example} that uses some of these techniques, and provide some concluding remarks in Sec.~\ref{sec:conclusions}. 

\section{Theory}
\label{sec:theory}
We use a simplified fluid model for gas discharges, based on the following equations
\begin{eqnarray}
  \label{eq:streamer_equations}
    &\partial_tn_i = \nabla\cdot\left(D_i\nabla n_i - \mathbf{v}_in_i\right) + S_i, \\
  &\nabla^2\phi = -\frac{\rho}{\epsilon_0}, \\
  \label{eq:rte}
    &\kappa_j\Psi_j - \nabla\cdot\left(\frac{1}{3\kappa_j}\nabla\Psi_j\right) = \frac{\eta_j}{c},
\end{eqnarray}
where $n_i$ is the density of species $i$, $D_i$ the diffusion coefficient, $\mathbf{v}_i$ the velocity, and $S_i$ the source term. The electric potential is given by $\phi$ and $\rho = \sum_{i}q_in_i$ is the space charge density. The symbols $\Psi_j$ denote the isotropic radiative density of a photon group $j$, $\kappa_j$ is the Beer's length and $\eta_j$ is an isotropic source term. 

Equations \eref{eq:streamer_equations}-\eref{eq:rte} are solved with finite volumes over an adaptive mesh based on Cartesian grid patches, and solid boundaries are treated with an embedded boundary formalism. The spatial discretization of Eqs.~\eref{eq:streamer_equations} allows for arbitrarily cut cells (even ones that contain multiple cell fragments), and is as follows:

\begin{enumerate}
\item The convective term $\nabla\cdot\left(\mathbf{v}_in_i\right)$ is discretized with the unsplit Godunov's method. The state at a face center is given by the solution to a Riemann problem with slope-limited left and right states; the Riemann solution is the upwind state at the face. One-sided slopes are used if there are not enough cells available for the left or right slopes, which can occur if a cell face lies completely inside a material. On cut-cells, we require the flux on face centroids in addition to face centers. This is done by first computing fluxes at face centers, and then interpolating these to the respective face centroids. We additionally stabilize the convective derivative by computing a hybrid divergence with charge redistribution that allows us to use a standard CFL condition for cut cells with arbitrarily small volume fractions. Charge injection into the domain is a part of the advective discretization, and thus the injected charge is also redistributed in the neighborhood of the cut cells. 

\item The diffusion advancement of $\partial_tn_i = \nabla\cdot\left(D_i \nabla n_i\right)$ is handled implicitly with the Twizell-Gumel-Arigu (TGA) scheme \cite{Twizell1996} (see equation~\eref{eq:tga}). This scheme is very stable in embedded boundary applications. The spatial discretization of the elliptic term $\nabla\cdot\left(D_i \nabla n_i\right)$ is done with a second-order accurate cell-centered discretization with Neumann boundary conditions everywhere. The resulting Helmholtz equations that arise from the TGA discretization are solved with a geometric multigrid method with red-black Gauss-Seidel smoothers and a biconjugate gradient stabilized method as the bottom solver. 

\item The remaining elliptic equations for the Poisson and the radiative transfer equations are discretized with second order accurate cell-centered solvers. The spatial discretization is the same as for the diffusion equation above, with the exception of boundary conditions. 
\end{enumerate}

The computer code that we use is compatible with adaptive mesh refinement and runs at high concurrencies (tested for up to 8k cores so far). A full discussion of the underlying software is not possible in the scope of this paper, but can be found elsewhere \cite{Marskar2018}. We will use this code to explore the numerical error obtained by using different-sized time steps for fluid and elliptic equations, thereby probing the importance of updating the Poisson and radiative transfer equations. We will then apply these findings to a large scale simulation example. 

\subsection{Regular time stepping}
\label{sec:rk2_tga}
First, we present a consistent second order time stepper that is suitable for embedded boundary applications, which we will refer to as the consistent scheme. This integrator is based on a second order monotone-in-time Runge-Kutta method together with second order implicit diffusion and advances $t^k \rightarrow t^{k+1}$ as follows:
\begin{enumerate}
\item Compute $n^\ast = n^k + \Delta t\left[S^k - \nabla\cdot\left(\mathbf{v}^kn^k\right)\right]$.
\item Compute $\mathbf{E}^\ast$ by solving the Poisson equation with the space charge density $\rho^\ast=\rho(n^\ast)$.
\item Compute radiative transfer source terms $\eta^\ast = \eta\left(n^\ast, \mathbf{E}^\ast\right)$ and obtain $\Psi_\gamma^\ast$ by solving the RTE equations.
\item Advance $n^\dagger = \frac{1}{2}\left(n^k + n^\ast + \Delta t\left[S^\ast - \nabla\cdot\left(\mathbf{v}^\ast n^\ast\right)\right]\right)$ where $S^\ast = S\left(n^\ast, \Psi^\ast, \bme^\ast \right)$ and $\mathbf{v}^\ast = \mathbf{v}\left(\bme^\ast\right)$
\item Compute $\mathbf{E}^\dagger$ by solving the Poisson equation with $\rho^\dagger = \rho\left(n^\dagger\right)$.
\item Compute diffusion coefficients $D = D\left(\mathbf{E}^\dagger\right)$ and obtain $n^{k+1}$ with an implicit diffusion advance
  \begin{equation}
    \label{eq:tga}
    \left(I-\mu_1L\right)\left(I - \mu_2L\right)n^{k+1} = \left(I+\mu_3L\right)n^\dagger,
  \end{equation}
  where $L$ is the diffusion operator. Expressions for $\mu_1$, $\mu_2$, and $\mu_3$ are found in \cite{Marskar2018}.
  For non-diffusive species then $n^\dagger\rightarrow n^{k+1}$.
\item Obtain the final electric field $\mathbf{E}^{k+1}$ by solving the Poisson equation with $\rho^{k+1} = \rho\left(n^{k+1}\right)$. 
\item Compute radiative transfer source terms $\eta^{k+1} = \eta\left(n^{k+1}, \mathbf{E}^{k+1}\right)$ and obtain $\Psi_\gamma^{k+1}$ by solving the RTE equations
\end{enumerate}
In the above, steps (i) through (iv) describe a consistent strongly stability preserving Runge-Kutta method of order two; steps (vi) and (vii) describe an implicit diffusion advance, and step (viii) describe the final radiative transfer update. For the physical model that we will consider, this scheme performs 11 elliptic updates per time step: Three Poisson updates, six radiative transfer equation solves, and two Helmholtz solves for the diffusion step. The computational bottlenecks are due to these elliptic updates. 

\subsection{Probing elliptic equations - a modified time stepper}
The modified scheme segregates the evolution into different-sized time steps. The outline of this method is as follows: We assume the existence of two disparate time scales $\Delta t_f$, which is a ''fine'' numerical time scale, and $\Delta t_c$ which is a ''coarse'' numerical time scale. We assume $\Delta t_f < \Delta t_c$ and $\Delta t_c = m\Delta t_f$ where $m > 0$ is an integer such that the time scales synchronize. Typically, $\Delta t_f$ is is smaller than the CFL time $\Delta t_{\cfl}$, whereas $\Delta t_c$ is a coarser numerical time scale, typically taken to be proportional to the dielectric relaxation time. The various time steps are summarized in Table~\ref{tab:timesteps}. We then evolve the equations of motions in the following way:
\begin{enumerate}
\item Evolve the advective-reactive part of the species equations from $t^k$ to $t^k + \Delta t_c$ by using $m$ non-diffusive steps with individual time step sizes $\Delta t_f < \Delta t_\cfl$, corresponding to steps (i)-(iv) in \Sref{sec:rk2_tga}. The advective discretization over a time step $\Delta t_f$ then occurs as follows:
  \begin{eqnarray}
    \label{eq:advec_react}
    n^\ast &= n^k + \Delta t_f\left[S^k - \nabla\cdot\left(\mathbf{v}^kn^k\right)\right], \\
    \label{eq:s_modified}
    S^\ast &= S\left(n^\ast, \Psi^k, \bme^k \right), \\
    \label{eq:v_modified}
    \bmv^\ast &= \bmv\left(\bme^k\right), \\
    n^\dagger &= \frac{1}{2}\left(n^k + n^\ast + \Delta t_f\left[S^\ast - \nabla\cdot\left(\mathbf{v}^\ast n^\ast\right)\right]\right).
  \end{eqnarray}
  This process is repeated $m$ times such that the equations are advanced a time $\Delta t_c$. 
\item Perform a diffusion advance $\left(I-\mu_1L\right)\left(I - \mu_2L\right)n^{k+1} = \left(I+\mu_3L\right)n^\dagger$. For non diffusive species, $n^\dagger \rightarrow n^{k+1}$.
\item Solve the Poisson equation with $\rho^{k+1} = \rho\left(n^{k+1}\right)$ and obtain the new electric field $\bme^{k+1}$.
\item Compute radiative transfer source terms $\eta^{k+1} = \eta\left(n^{k+1}, \mathbf{E}^{k+1}\right)$ and obtain $\Psi_\gamma^{k+1}$ by solving the RTE equations
\end{enumerate}
We will examine three cases in total so that we can estimate which elliptic equations are important for numerical accuracy, and which ones are not: 
\begin{description}
\item{Case 1:} Update all elliptic equations at the coarse time step $\Delta t_c$ as above. 
\item{Case 2:} Also update $\bme$ consistently. In this case, we additionally update the electric field at each Runge-Kutta stage of \eref{eq:advec_react}. 
\item{Case 3:} Additionally update the RTE equations consistently. In this case, we additionally update the radiative transfer equations at each Runge-Kutta stage of \eref{eq:advec_react}.
\end{description}
Thus, the electric field and radiative transport equations are updated at the either the coarse or the fine numerical time step, allowing us to assess the importance of both types of elliptic updates. Note that, at worst, the above scheme is first order accurate in $\Delta t_c$ due the modified electric field coupling. 

\begin{table}[h!t!b!]
  \begin{tabular}{lll}
    Symbol & Formula & Description \\
    \hline
    $\Delta t_c$ & - & Coarse numerical time step \\
    $\Delta t_f$ & - & Fine numerical time step \\
    $\Delta t_{\bme}$ & $\epsilon_0\left|\bme\right|/\mathbf{J}$ & Dielectric relaxation time \\
    $\Delta t_{\cfl}$ & $\Delta x/\left|\mathbf{v}\right|$ & CFL time step \\
    $\Delta t_{S}$ & $S/n$ &Ionization time step \\
  \end{tabular}
  \caption{Time step size nomenclature. The time steps $\Delta t_{\cfl}$ and $\Delta t_{S}$ are minimized according to the worst offender (usually electrons). }
  \label{tab:timesteps}
\end{table}

The rationale for the above approach is removal of potentially redundant elliptic solves at every $\Delta t_f$, which would be present for a consistent scheme. Indeed, one should not need to update the Poisson equation if $\Delta t_f \ll \Delta t_{\bme}$, but rather resolve its physical time scale $\Delta t_\bme$ at reasonable accuracy. It is quite clear that an electric field update is redundant when space charge currents are negligible, which is e.g. the case in the avalanche stage of a streamer discharge. Elliptic solves are bottlenecks for most streamer simulation codes, and unnecessary updates can easily become performance killers. This is particularly true when direct solvers are involved, but also for iterative solvers when curved solid boundaries are present since this leads to deteriorated convergence rates for multigrid smoothing. One may possible extend these ideas to kinetic models, although the accuracy of this simplification must then be determined by a kinetic code, which is not a topic in this paper. 

The use of longer time steps for the Poisson equation is not new. In the PIC code in \cite{Teunissen2016}, the authors update the electric field using a time step that may be as long as $\Delta t_{\bme}$, i.e. $\Delta t_c \leq \Delta t_{\bme}$. Likewise, the authors in \cite{Fierro2014} use a fixed time step for the Poisson update and a much finer time step for the particle push. The accuracy of this approximation was not quantified in either paper. Relatedly, the authors of the fluid code in \cite{Teunissen2017} and the authors of the PIC code in \cite{Fierro2018} use a much stricter condition, updating the Poisson equation consistently at each advective step or particle push. For the conditions in \cite{Fierro2018}, the time steps are on the order of $1\,\fs$ whereas the dielectric relaxation time is estimated to be on the order of $100\,\fs$. On the other hand, for most fluid simulations reported to date, the electric field updates are consistent, i.e. they are performed at each advective or diffusive step, regardless of the dielectric relaxation time.

\section{Two-dimensional simulation experiments}
\label{sec:2D}
In this section we estimate how often the $\bme$-field should be updated by means of two-dimensional numerical experiments. We consider a Cartesian $(2\,\cm)^2$ domain with a rod-plane geometry. The rod is a cylinder with a hemispherical cap with a radius of $500\,\um$ protruding $1\,\cm$ from the center of the top domain edge. This edge, and the needle, is live with a voltage of $40\,\kV$ whereas the bottom edge is grounded. For simplicity, we use homogeneous Neumann boundary conditions on the left and right side edges.

For chemistry, we consider a three-species model for air at 1 bar that consists of electrons $n_\electron$, positive ions $n_+$, and negative ions $n_-$. The kinetic coefficients for this scheme are given in \cite{Morrow1997}. All three species are advected, but only electrons are diffusive. Note that for positive streamer discharges that do not touch the cathode, one might possibly simplify the system further by assuming immobile ions, although this is not possible for negative streamer due to the dynamics of the cathode sheath. For radiative transport, we consider the model by \cite{Segur2006, Bourdon2007}. The initial conditions are $n_\electron = n_+ = 10^{10}\,\m^3$ and the equations are integrated for $2\,\ns$, which is sufficiently long for capturing the three basic simulations phases: electron avalanche, streamer inception, and streamer propagation phase. 

The numerical error of the modified scheme is evaluated by comparison with the consistent scheme at each step. Let $n^{k+1}$ denote the final state after advancing the species equation from $t_k$ to $t_k + \Delta t_c$ by using the modified scheme. Let $\widetilde{n}^{k+1}$ denote the final state after advancing the species equation from $t_k$ to $t_k+\Delta t_c$ by using the consistent scheme. The relative error obtained for $n^{k+1}$ at each grid point $\bmi$ is then
\begin{equation}
  \label{eq:error}
  \triangle n_{\bmi}^{k+1} = n_{\bmi}^{k+1} - \widetilde{n}_{\bmi}^{k+1}.
\end{equation}
We compute the $L_2$ error norm as
\begin{equation}
  L_2\left(\triangle n^{k+1}\right) = \sqrt{\frac{\sum_{\bmi} \kappa_{\bmi}\left(\triangle n_{\bmi}^{k+1}\right)^2}{\sum_{\bmi} \kappa_{\bmi}\left(\widetilde{n}_{\bmi}^{k+1}\right)^2}},
\end{equation}
where $\kappa_{\bmi}$ is the volume fraction of a cut-cell (see \cite{Marskar2018} for details). We remark that $\triangle n^k$ is the step-wise error caused by inconsistent elliptic updates and not the accumulated error. 

\begin{figure*}[h!t!b!]
  \centering
  \includegraphics[height=.8\textheight]{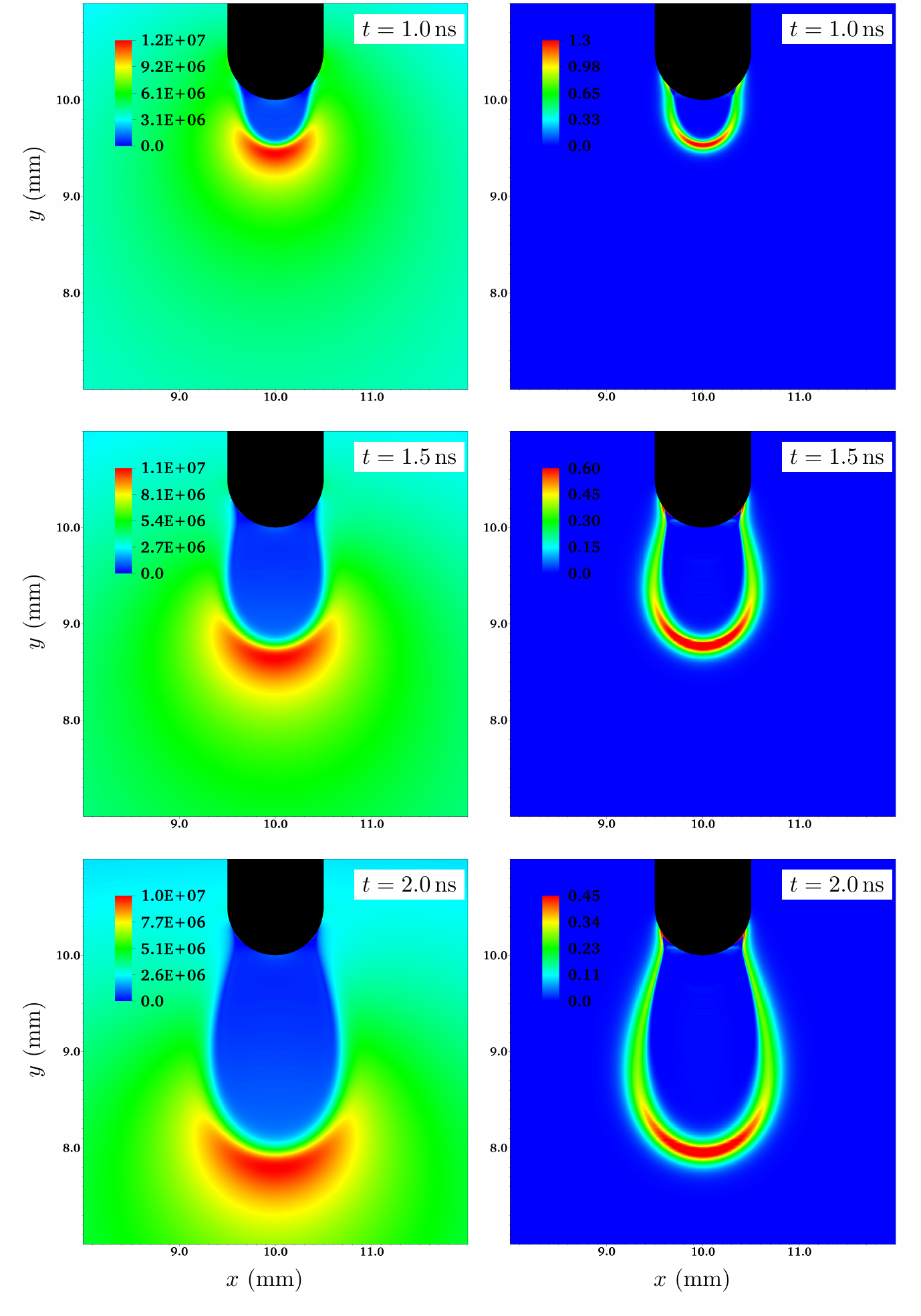}
  \caption{Two-dimensional streamer evolution with $\Delta t_c = \min\left(\Delta t_{\bme}, \Delta t_{S}\right)$ with radiative transport updates at $\Delta t_c$. The left hand side column shows the electric field magnitude $\left|\bme\right|$ and the right hand side column shows the space charge density $\rho$. All numbers are in SI units and times are indicated in each frame. }
  \label{fig:evolution_2d}
\end{figure*}

We use a fine spatial grid for all the simulations presented in this section. The grids use a base mesh of $(128)^2$ cells and includes four refinement levels; the refinement factor between each level is $4$ except for the last one where it is $2$. This yields an effective domain of $(16384)^2$, corresponding to a finest-level resolution of $1.2\,\um$.

\Fref{fig:evolution_2d} shows the time-evolution of a simulation that updates the elliptic equations at $\Delta t_c$ and restricts the time step according to
\begin{equation}
  \Delta t_c = \min\left(\Delta t_{\bme}, \Delta t_{S}\right),
\end{equation}
where the fine step restriction is $\Delta t_f \leq 0.5\Delta t_\cfl$. The equations are integrated for $2\,\ns$ and evolve without obvious numerical errors such as spurious oscillations or numerical instabilities. This is true even though the number of advective integration steps per elliptic update ranged up to thirty, with an average of twenty advective steps per elliptic update throughout the simulation. 

\begin{figure*}[h!t!b!]
  \centering
  \includegraphics{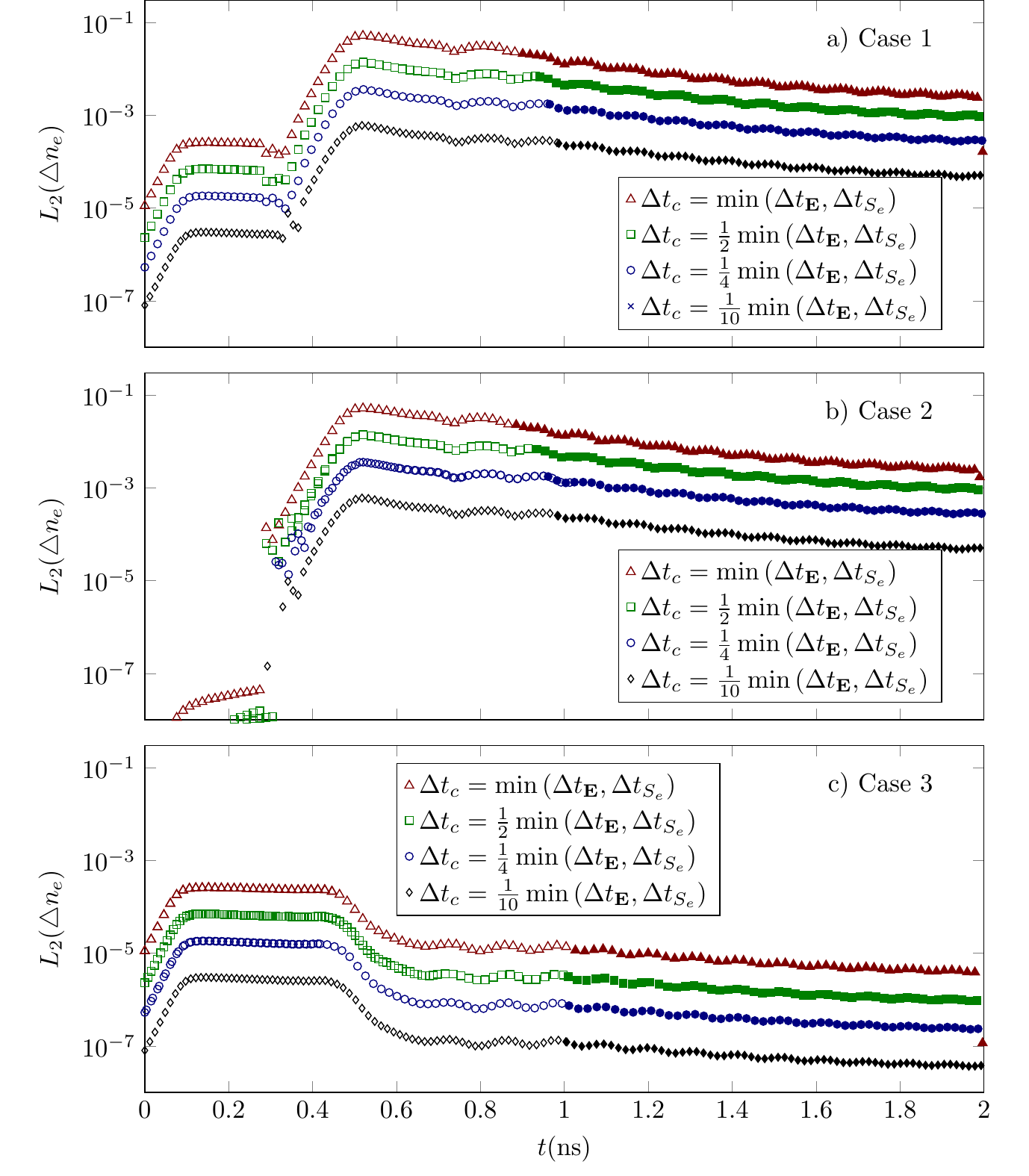}
  \caption{Error estimators for two-dimensional streamer simulation experiments. The various symbols show different time stepping criteria: If a symbol is empty (filled), the time step was restricted by $\Delta t_{S_e}$ ($\Delta t_{\bme}$). The solid line shows the numerical error for a fully consistent scheme. a) Poisson and radiative transfer equations are updated at $\Delta t_c$. b) Radiative transport updates at $\Delta t_f$. c) Poisson updates at $\Delta t_f$.}
  \label{fig:convergence_2d}
\end{figure*}

Next, to answer the question of when we need to update the electric field and radiative transport equations, we perform additional simulations by using smaller values of $\Delta t_c$ for each of the three cases listed in the preceding section. Specifically, we take $\Delta t_c = \alpha\min\left(\Delta t_{\bme}, \Delta t_{S_e}\right)$, where $\alpha = 0.1, 0.25, 0.5, 1.0$, and examine the numerical error for each case. The results of these comparisons are shown in \fref{fig:convergence_2d} for the electrons $n_e$. To aid in the interpretation of this error diagram, we mention that the error rise around $t=0.3\,\ns$ coincides with the inception of the streamer; whereas the initial propagation phase takes place from $0.5\,\ns$ and beyond. The various figures show the errors for the three different cases above. \Fref{fig:convergence_2d}a) shows the error norms when all elliptic equations are updated at $\Delta t_c$, corresponding to Case 1. All simulations are stable, although the numerical error can be significant in certain parts of the streamer evolution. For the propagation phase of the streamer, the error decreases to about $10^{-3}$, which is (in our opinion) a more acceptable error. For the avalanche phase, which occurs for $t<0.3\,\ns$ in \fref{fig:convergence_2d}, the obtained errors are due to absence of consistently updated photoionization. However, the computed errors are smaller than $2\times10^{-4}$. \Fref{fig:convergence_2d}b) shows Case 2 where we consistently update the radiative transport equations. Differences between \fref{fig:convergence_2d}a) and \fref{fig:convergence_2d}b) are only seen in the avalanche phase $t<0.3\,\ns$. In this evolution regime we have $\Delta t_{\cfl} \ll \Delta t_{\bme}$ so that electric field updates are redundant; solution errors occur due to consistent radiative transport updates, but even then the errors are quite small. For the streamer inception and propagation phases $t>0.3\,\ns$, the error increases for both \fref{fig:convergence_2d}a) and b), but there is no notable quantitative difference between the computed errors. This error must therefore be due to inconsistent electric field updates. \Fref{fig:convergence_2d}c) correponds to Case 3 above, where radiative transport is updated inconsistently and the electric field is updated consistently. The initial errors for the avalanche phase $t<0.3\,\ns$ correpond to the errors in \fref{fig:convergence_2d}a) and are due to inconsistent photoionization updates. For Case 3 the error is small throughout the entire evolution, showing that inconsistent photoionization updates leads to minor accuracy losses. 

\begin{figure*}[h!t!b!]
  \centering
  \includegraphics{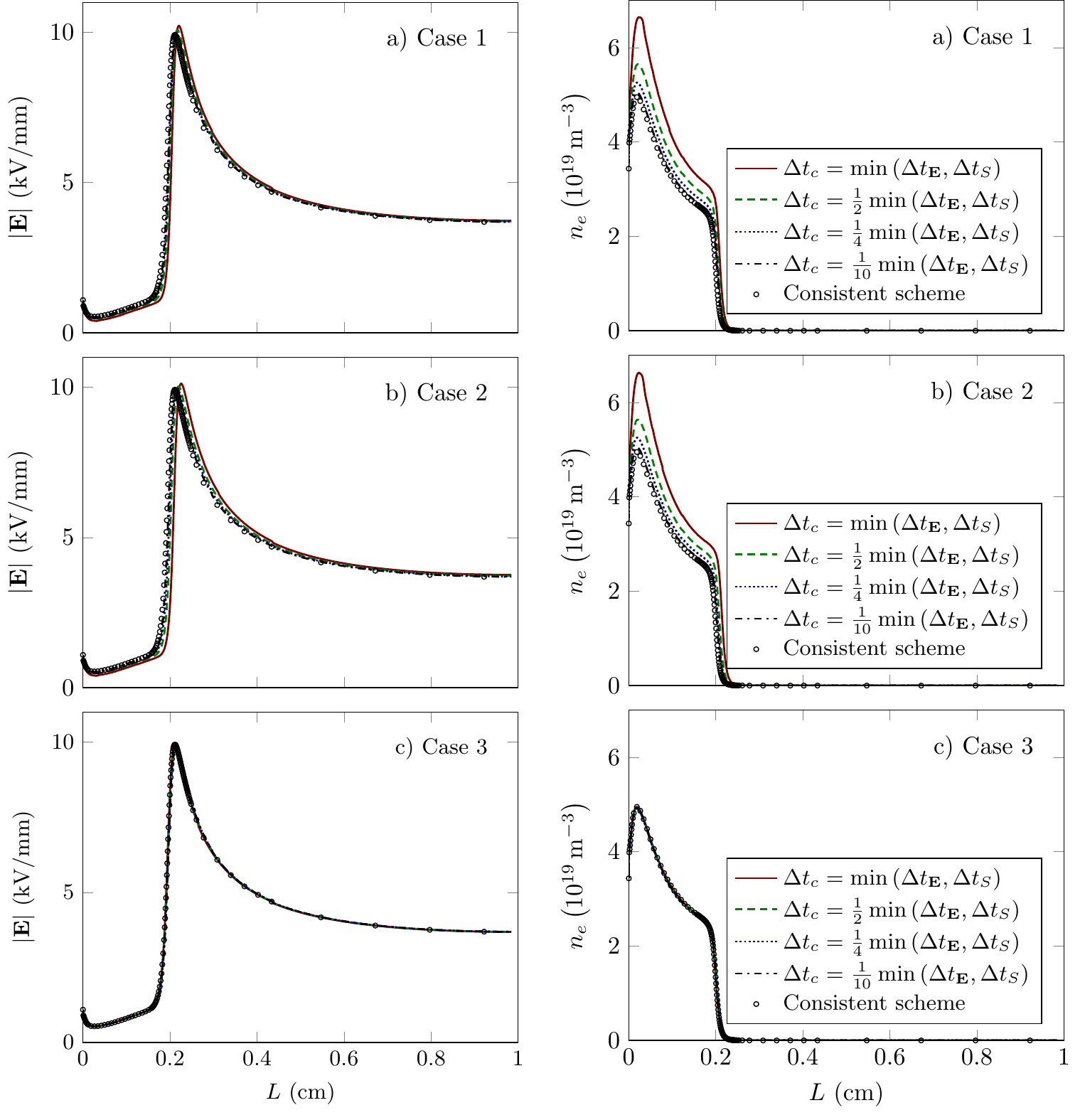}
  \caption{Plots of the electric field magnitude and electron densities on the symmetry axis after integrating the equations for $2\,\ns$ - corresponding to \fref{fig:convergence_2d}. The various rows show: a) Poisson and radiative transfer equations are updated at $\Delta t_c$. b) Radiative transport updates at $\Delta t_f$. c) Poisson updates at $\Delta t_f$. }
  \label{fig:field_comparison}
\end{figure*}

To show the effect of the accumulated errors on the streamer for each of the nine simulations above, \fref{fig:field_comparison} shows the electric field and electron profiles on the symmetry axis at the end of the simulations. As before, \Fref{fig:field_comparison}a) shows Case 1, \fref{fig:field_comparison}b) shows Case 2, and \fref{fig:field_comparison}c) shows Case 3. We only find minor differences with respect to the initial velocities and peak field amplitude of the streamers. For the electron densities, \fref{fig:field_comparison} shows that omission of electric field updates may lead to significant variation of the predicted electron densities. For the worst resolved case in \fref{fig:field_comparison}a) which updates the electric field at the relaxation time, the predicted peak electron density is about $25\%$ higher than it should be. Likewise, \fref{fig:field_comparison}c) shows again that consistent updates of the radiative transport model do not lead to improved model accuracy.

\begin{table}[h!t!b!]
  \centering
  \begin{tabular}{c|c|c|c}
    $\Delta t_c$ & Final $L_2(n_e)$ & \begin{tabular}{c} Poisson updates \\ (modified scheme) \end{tabular} & \begin{tabular}{c} Poisson updates \\ (original scheme) \end{tabular} \\
    \hline
    $\textrm{min}\left(\Delta t_{\bme}, \Delta t_S\right)$ & $2.43\times 10^{-1}$ & 133 &  8745 \\
    $\frac{1}{2}\textrm{min}\left(\Delta t_{\bme}, \Delta t_S\right)$ & $1.10\times 10^{-1}$ & 237 & 8862 \\
    $\frac{1}{4}\textrm{min}\left(\Delta t_{\bme}, \Delta t_S\right)$ & $4.09\times 10^{-2}$ & 450 & 9216 \\
    $\frac{1}{10}\textrm{min}\left(\Delta t_{\bme}, \Delta t_S\right)$ & $2.72\times 10^{-2}$ & 1095 & 9966 
  \end{tabular}
  \caption{Total number of Poisson updates for the inconsistent (Case 1) and consistent schemes. }
  \label{tab:elliptic_solves}
\end{table}

In summary, the preceding discussion shows that for the radiative transfer model considered here one does not need to compute the radiative transfer equations with a fine temporal resolution. Updates of the electric field should be evaluated based on a tradeoff between simulation time and accuracy need. For the model considered here, updating the electric field at $\Delta t_{\bme}/10$ yields acceptable accuracy (a maximum step error of $10^{-3}$). We have also computed the final $L_2$ error norm for the electron density by computing error norms between the inconsistent (Case 1) solution and the consistent solution. These accumulated errors are reported in \tref{tab:elliptic_solves} where we also show the reduction of the number of elliptic solves for the inconsistent scheme. When we update the electric field at $\Delta t_{\bme}/10$, the number of Poisson updates is reduced by approximately a factor nine, and the number of radiative transfer updates are reduced by a factor of eighteen. The price to pay for this is a $2.7\%$ increased error in the $L_2$ norm. 

\section{A high performance computing example}
\label{sec:hpc_example}
\begin{figure*}[h!t!]
  \centering
  \includegraphics[width=0.6\textwidth]{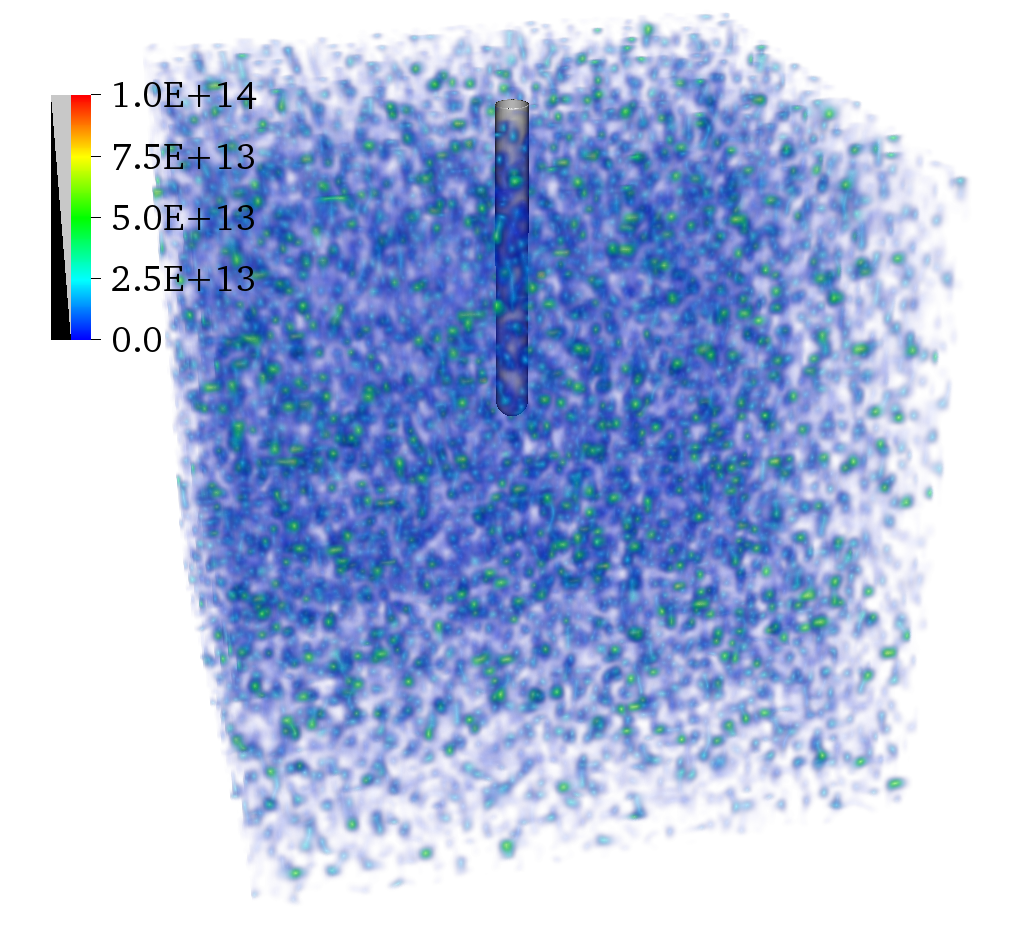}
  \caption{Initial preionization (in SI units) for the example simulation.}
  \label{fig:electron_initial}
\end{figure*}

As a large-scale computational example that uses some of these techniques, we next consider the inception and propagation of streamers in a needle-plane geometry. The domain is a $(2\,\cm)^3$ cube with a needle electrode with a $500\,\um$ radius protruding $1\,\cm$ from the center of the top domain face (see \fref{fig:electron_initial}). This face, and the needle, is live with a voltage of $15\,\kV$ whereas the bottom face is grounded. Homogeneous Neumann boundary conditions are used for the Poisson equation on the remaining domain faces. To demonstrate the use of spatially adaptive methods to streamers, we will consider initial conditions that provoke streamer branching and prevent prediction of their paths. Streamers are initiated by considering a stochastic preionization level with peak amplitudes up to $10^{14}\,\m^{-3}$ where the initial density is prescribed by using a landscape function borrowed from computer graphics \cite{Perlin2002} which guarantees that the resulting noise is $C^1$ smooth (we use the original hash table in \cite{Perlin2002}). This function is then exponentiated in order to generate randomly placed plasma spots. The correlation length between these spots is roughly $500\,\um$. The initial electron distribution is shown in \fref{fig:electron_initial}. 

The simulation was run with a maximum time step $\Delta t_c = \textrm{Min}\left(\Delta t_\bme, \Delta t_S\right)$ until $t=15\,\ns$ with a time-to-solution of approximately \simtime\,hours. The electric field was updated at every advective evaluation whereas radiative transfer was updated at $\Delta t_c$ only (corresponding to Case 3 above). The simulation was run on 128 computing nodes interconnected with Infiniband. Each node contains dual-socket Intel Broadwell (E5-2683v4) chips with 32 cores per node in total. MPI ranks were mapped to cores with a one-to-one ratio for a total concurrency of 4096. For grids, we use a patch-based AMR grid generated by the Berger-Rigoutsos algorithm\cite{Berger1991} with a blocking factor of $16$ and a maximum patch size of $32$. This algorithm takes as input a number of ''tags'', which specify which cells should be refined, and the output of this algorithm is a properly nested hierarchy of Cartesian grids. The domain is discretized using this algorithm on a base mesh of $(256)^3$ cells with five levels of mesh refinement, which yields an effective domain of $(8192)^3$, corresponding to an effective resolution of $2.44\,\um$. This resolution is sufficient for numerical stability, but we remark that we have not performed grid convergence studies for this example. The refinement criteria that we use is based on resolving streamer heads and space charge layers \cite{Marskar2018}. The time step restriction on $\Delta t_f$ is done with $\Delta t_f \leq 0.5\Delta t_{\cfl}$. The number of $\Delta t_f$ steps per coarse step $\Delta t_c$ was between $1$ and $20$, with an average of $2$ throughout the simulation. For an even finer spatial resolution, this ratio would be even higher, whereas this ratio would be 1:1 for a fully consistent scheme, which would increase the simulation time by approximately $40\%$. 

\begin{figure*}[t]
  \centering
  \includegraphics[width=.45\textwidth]{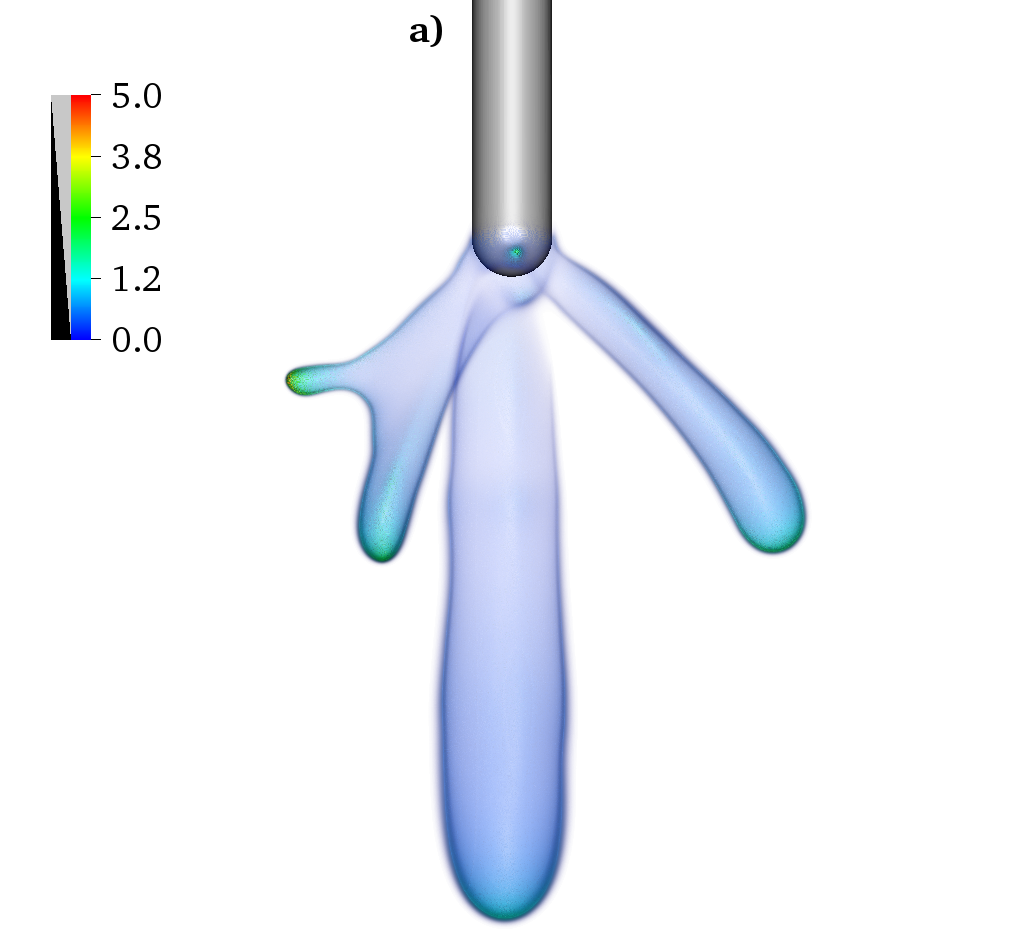}
  \includegraphics[width=.45\textwidth]{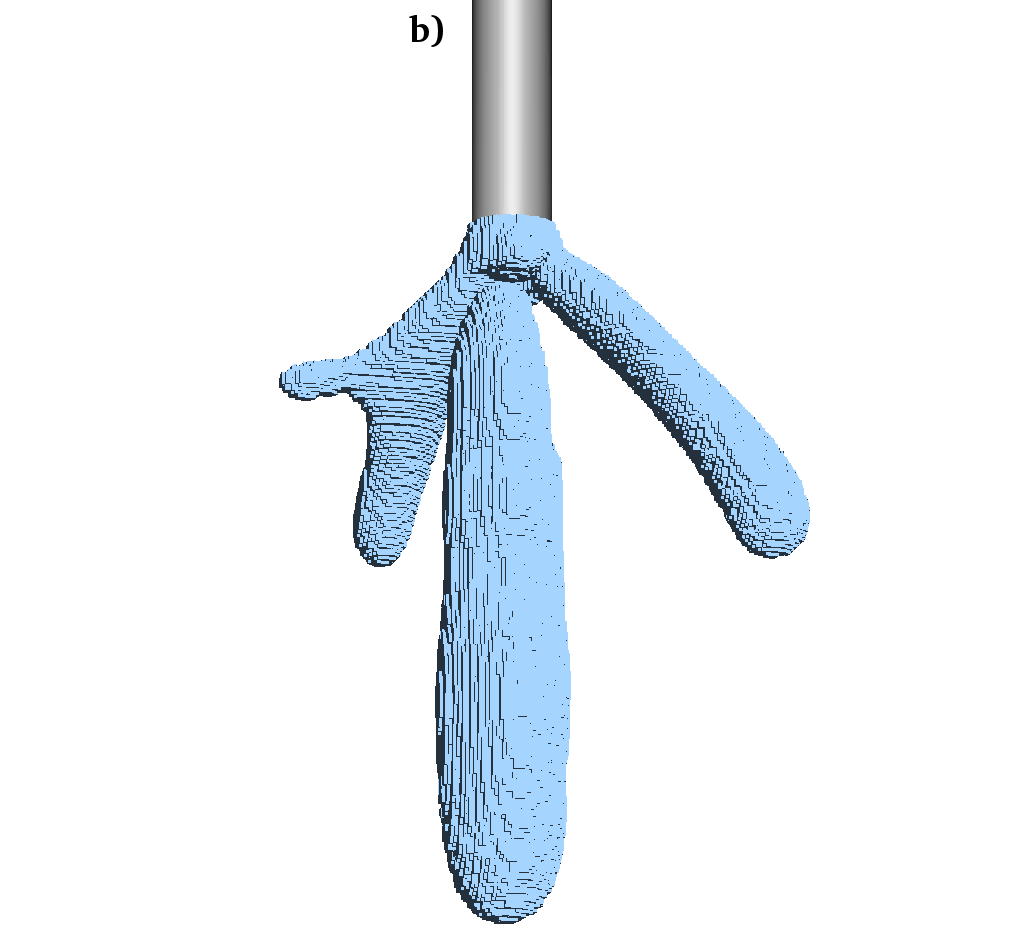} \\[2em]
  \includegraphics[width=.45\textwidth]{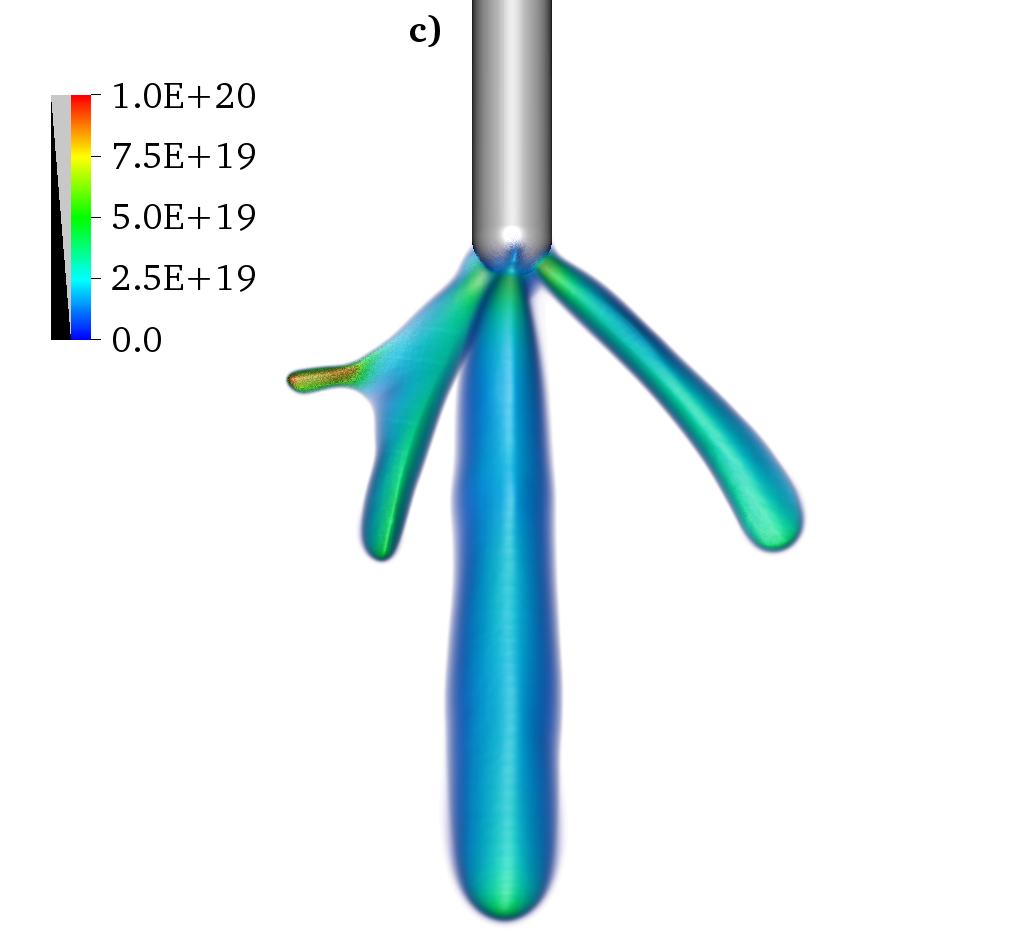}
  \includegraphics[width=.45\textwidth]{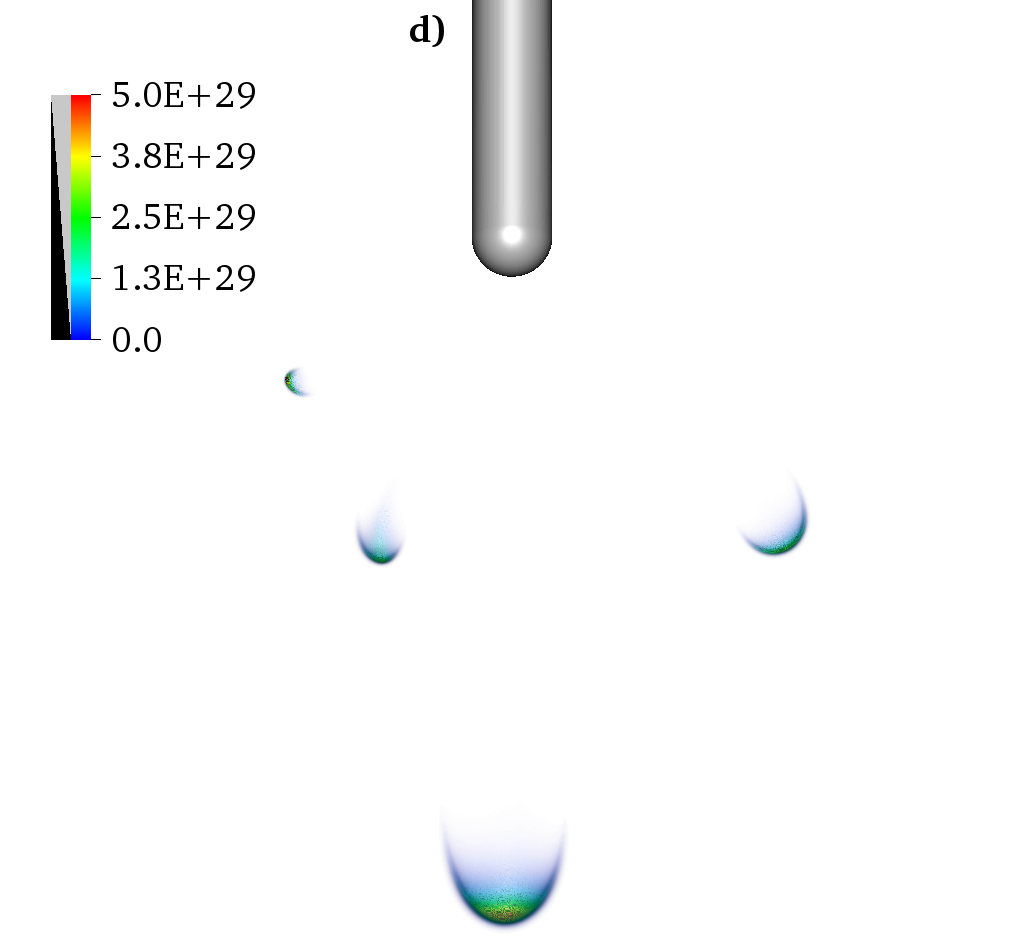}
  \caption{Final simulation state after $15\,\ns$. a) Volume rendered space charge density. The data range has been adjusted to enhance visibility. b) Patch distribution at the finest level. Each brick represents a computational unit of minimum size $(16)^3$ and maximum size $(32)^3$. The ''inside'' of the brick agglomeration is empty (not shown). c) Volume rendered electron density. d) Volume rendered electron source term. All quantities are presented in SI units.}
  \label{fig:solution}
\end{figure*}

\Fref{fig:solution} shows the final simulation state after $15\,\ns$. The various subplots in the figure show snapshots of the electron density, space charge density, electron source term, and the mesh distribution on the finest level. Due to the perturbed initial conditions, we find that multiple streamers start from the electrode. During inception, we do not find uniform field screening over the needle tip (not shown) although this is something that we observe in simulations that do not use perturbed initial conditions. For this simulation, we observe five initial streamers, but two of these start from a position slightly higher up on the electrode and stop after a few nanoseconds. The other three filaments propagate into the gap. Their cross sections are generally not circular; for one of the branches the ratio between the major and minor radii is roughly $2$ so that the cross section is comparatively flat, which we believe is due to electrostatic repulsion from the other two nearby branches. Furthermore, this filament branches after approximately $10\,\ns$, as seen in \fref{fig:solution}. Towards the end of the simulation, four individual streamer heads that propagate in different directions can clearly be observed, see e.g. \fref{fig:solution}d). The largest streamer propagates almost parallel with the rod, whereas the smallest streamer (see the branch in Fig.~\ref{fig:solution}) propagates almost perpendicular with it. 

\section{Conclusion}
\label{sec:conclusions}
We have discussed recent advances in multiresolution computer models for streamer discharges. The use of longer time steps for the electric field and radiative transport leads to inherent numerical errors, which we quantify by means of two-dimensional simulations. We show that it is permissible to use a longer time step for the elliptic equations (e.g. Poisson) than for advection and chemistry, and believe that such techniques can be used to speed up simulation cases. The ideas can possibly also be extended to kinetic simulations. A high-performance computing example that uses some of these techniques was then presented.

\section*{Acknowledgements}
This work was financially supported by the Research Council of Norway through project 245422 and industrial partner ABB AS, Norway. The computations were performed on resources provided by UNINETT Sigma2 - the National Infrastructure for High Performance Computing and Data Storage in Norway.

\bibliography{references}

\end{document}